\documentclass[fleqn,usenatbib]{mnras}

\usepackage{amssymb}	
\usepackage{booktabs,tabularx,graphicx,amsmath,makecell,hhline,enumitem,gensymb,multirow,ulem,pifont,adjustbox}
\usepackage{hyperref} 
\usepackage[dvipsnames]{xcolor}
\usepackage{newtxtext,newtxmath}

\usepackage[T1]{fontenc}

\DeclareRobustCommand{\VAN}[3]{#2}
\let\VANthebibliography\thebibliography
\def\thebibliography{\DeclareRobustCommand{\VAN}[3]{##3}\VANthebibliography}

\usepackage{graphicx}	
\usepackage{amsmath}	

\title[Two massive early accretion events in an MW-like galaxy]{The impact of two massive early accretion events in a Milky Way-like galaxy: repercussions for the buildup of the stellar disc and halo}

\author[M. D. A. Orkney et al.]{Matthew D. A. Orkney$^{1}$\thanks{E-mail: morkney@icc.ub.edu},
Chervin F. P. Laporte$^{1}$, Robert J. J. Grand$^{2,3}$, Facundo A. G\'omez$^{4}$, \newauthor Freeke van de Voort$^{5}$, Federico Marinacci$^{6}$, Francesca Fragkoudi$^{7}$, Ruediger Pakmor$^{8}$, Volker Springel$^{8}$ \\
$^{1}$ICCUB, Universitat de Barcelona, Mart\'{i} i Franqu\`{e}s 1, 08028 Barcelona, Spain\\
$^{2}$Instituto de Astrof\'isica de Canarias,
Calle Via Lactea s/n, E-38205 La Laguna, Tenerife, Spain\\
$^{3}$Departamento de Astrof\'isica, Universidad de La Laguna,
Av. del Astrof\'isico Francisco S\'anchez s/n, E-38206, La Laguna, Tenerife, Spain\\
$^{4}$Departamento de F\'isica y Astronom\'ia, Universidad de La Serena, Av.
Juan Cisternas 1200 Norte, La Serena, Chile\\
$^{5}$Cardiff Hub for for Astrophysics Research and Technology, School of Physics and Astronomy, Cardiff University, Queen's Buildings, The Parade, Cardiff\\ CF24 3AA, UK \\
$^6$Department of Physics \& Astronomy ``Augusto Righi'', University of Bologna, via Gobetti 93/2, 40129 Bologna, Italy\\
$^{7}$Institute for Computational Cosmology, Department of Physics, Durham University, South Road, Durham DH1 3LE, UK \\
$^{8}$Max Planck Institut f{\"u}r Astrophysik, Karl-Schwarzschild-Straße 1, 85748 Garching bei München, Germany  \\
}

\date{Accepted XXX. Received YYY; in original form ZZZ}

\pubyear{2022}

\begin{document}
\label{firstpage}
\pagerange{\pageref{firstpage}--\pageref{lastpage}}
\maketitle

\begin{abstract}
We identify and characterise a Milky Way-like realisation from the Auriga simulations with two consecutive massive mergers $\sim2\,$Gyr apart at high redshift, comparable to the reported Kraken and Gaia-Sausage-Enceladus. The Kraken-like merger ($z=1.6$, $M_{\rm Tot}=8\times10^{10}\, \rm{M_{\odot}}$) is gas-rich, deposits most of its mass in the inner $10\,$kpc, and is largely isotropic. The Sausage-like merger ($z=1.14$, $M_{\rm Tot}=1\times10^{11}\, \rm{M_{\odot}}$) leaves a more extended mass distribution at higher energies, and has a radially anisotropic distribution.
For the higher-redshift merger, the stellar mass ratio of the satellite to host galaxy is high (1:3). As a result, the chemistry of the remnant is indistinguishable from contemporaneous in-situ populations, making it challenging to identify through chemical abundances. This naturally explains why all abundance patterns attributed so far to Kraken are in fact fully consistent with the metal-poor in-situ so-called \textit{Aurora} population and thick disc. However, our model makes a falsifiable prediction: if the Milky Way underwent a gas-rich double merger at high redshift, then this should be imprinted on its star formation history with bursts about $\sim2\,$Gyrs apart. 
This may offer constraining power on the highest-redshift massive mergers.
\end{abstract}




\section{Introduction}

The assembly of galaxies in the $\Lambda$ Cold Dark Matter cosmology is predicted to progress via the gravitational collapse and hierarchical merging of dark matter haloes \citep{white1991}, which are themselves funnelled along a cosmic web of interconnecting sheets and filaments. This `accretion history' has considerable consequences for the evolution and present-day properties of galaxies \citep{kauffmann1993,moster2018}, highlighting the importance of simulating galaxies within a full cosmological context. The dynamical timescales within stellar haloes can be of order $\mathcal{O}$(Gyr), and the chemical evolution of its various merging structures depends on their mass. Therefore, it is possible to dissect a galaxy's assembly through careful analysis of its present-day phase-space and chemistry. \par

There is growing evidence that the Milky Way (MW) has itself been subjected to an eventful accretion history \citep{bell2008, helmi1999, deason2013, malhan2022}. Chemo-dynamical analysis of data from the \textit{Gaia} satellite \citep{GaiaDR1} has revealed that the inner halo of the MW is dominated by a massive radially anisotropic component \citep{belokurov2018, helmi2018}, confirming earlier observational evidence using kinematic and chemical data (metallicity and $\alpha$-abundances) for halo stars in the solar neighbourhood \citep[e.g.][]{chiba2000, brook2003, meza2005}. This is believed to be the remains of the so-called Gaia Sausage Enceladus (GSE), named due to its characteristic shape when viewed in velocity space. Inferences from observed chemistry \citep[e.g.][]{helmi2018} and cosmological simulations \citep[e.g.][]{belokurov2018,mackereth2019} suggest the GSE accreted $\approx10\,$Gyr ago, and was a massive merger with a mass between that of the Small and Large Magellanic Clouds. \par

There have now been suggestions of an even more ancient accretion event, termed `Kraken', from age-metallicity data of MW globular clusters \citep{Kruijssen2019, massari2019}. The stellar debris that has been attributed to Kraken inhabits the inner few kpc of the MW halo, right at the bottom of the gravitational potential well and argued to be chemically distinct, with Kraken stars exhibiting chemistry consistent with an early disrupted dwarf \citep{horta2021,Naidu2022b}. \par

Whether the MW underwent such a double merger remains highly uncertain, and little theoretical work has been carried out to study the impact and predictions of a double merger on the stellar halo or disc in light of the currently available observational constraints.

Here, we focus on a simulated MW-like galaxy with an early accretion history dominated by two massive mergers. The galaxies involved appear qualitatively similar to the proposed Kraken and GSE of the MW. In Section \ref{sec:method}, we introduce the Auriga project.  In Section \ref{sec:results}, we study the impact of both satellites on the stellar halo in chemodynamical space. In Section \ref{sec:discussion}, we discuss how the predictions from our identified Kraken-like object show qualitative trends that are consistent with the newly uncovered \textit{Aurora} in-situ disc population \citep{belokurov2022} and why identifying Kraken through simple chemical cuts is proving more difficult than previously thought, and will need to be tested with other falsifiable predictions. We conclude in Section \ref{sec:conclusion}. \par

\begin{table}
\setlength{\tabcolsep}{6pt} 
\centering
\resizebox{\columnwidth}{!}{
\begin{tabular}{lcc}
\toprule
\textbf{Property} & \textbf{Merger~1 (Kraken-like)} & \textbf{Merger~2 (GSE-like)} \\
\midrule
$z_{\rm infall, merge}$ & 2.15, 1.62 & 1.34, 1.14 \\
Merger ratio & 1:3 & 1:9 \\
Gas fraction & 0.76 & 0.85 \\
Peak $M_{\rm *}$ [M$_{\odot}$] & $1.46\times10^{9}$ & $2.71\times10^{9}$ \\
Peak $M_{\rm Tot}$ [M$_{\odot}$] & $8.06\times10^{10}$ & $1.11\times10^{11}$ \\
\text{[Fe/H]}, $\sigma$[Fe/H] & $-1.19,0.67$ & $-1.23,0.52$ \\
\text{[Mg/Fe]}, $\sigma$[Mg/Fe] & $0.23,0.07$ & $0.21,0.06$ \\
\bottomrule
\end{tabular}
}
\caption{Key properties for the two identified mergers. We define $z_{\rm infall}$ as the time when the merging halo crosses the $R_{200}$ radius of the host halo, and $z_{\rm merge}$ as the time when the satellites can no longer be tracked by {\sc subfind}. The merger ratios (stellar mass) are determined at $z_{\rm infall}$. Gas fractions are calculated at infall as $M_{\rm gas}/(M_{\rm gas} + M_*)$, within the stellar half-mass radius. Metallicities are determined for pre-infall member stars at $z=0$.}
\label{tab:merger_comparison}
\end{table}

\section{Methods} \label{sec:method}

The Auriga project consists of thirty cosmological magneto-hydrodynamical simulations of MW-like galaxies, with virial masses in the range $1-2\times10^{12}\,\text{M}_{\odot}$ \citep{Auriga}. They have been shown to have realistic properties that are broadly consistent with those of the MW \citep{monachesi2019, gomez2017, Auriga, fragkoudi2020}. The simulations were run with the Tree-PM moving-mesh code {\sc arepo} \citep{Arepo} using the `zoom' approach \citep{katz1993}. Auriga assumes cosmological parameters from \citet{Planck2014}, which are $\Omega_{\rm m}=0.307$, $\Omega_{\rm b}=0.04825$, $\Omega_{\rm \Lambda}=0.693$ and a Hubble constant of $H_0=100h\,\text{km}^{-1}\,\text{Mpc}^{-1}$, where $h=0.6777$. The zoom-in region of the target halo is resolved with a dark matter particle mass of $\sim3\times10^5\,\text{M}_{\odot}$ and baryonic mass of $\sim5\times10^4\,\text{M}_{\odot}$. Auriga includes physical models for a spatially uniform photoionising UV background, primordial and metal line cooling, star formation, stellar evolution and stellar feedback, supermassive black hole growth and feedback, and magnetic fields. We refer the reader to \citet{Auriga} for a full description. \par

In \citet{fattahi2019}, ten of the Auriga galaxies are identified as possessing features resembling the GSE in spherical $v_r, v_{\phi}$ coordinates. In this Letter we highlight one of these ten galaxies, Auriga 24 (Au-24), for which the early accretion history is dominated by two massive mergers occurring at $z=1.62$ and $z=1.14$. 

Our definition of merger stars includes all stellar particles which were at any time bound to the descendants of the merging galaxy at its peak mass, and which formed prior to the first pericentre passage around the host galaxy. Unless otherwise stated, galaxy properties are calculated for all bound simulation particles. \par

\section{Results} \label{sec:results}

\begin{figure}
\centering
  \setlength\tabcolsep{2pt}%
    \includegraphics[keepaspectratio, trim={0.2cm 0.1cm 0.2cm 0.1cm}, width=\columnwidth]{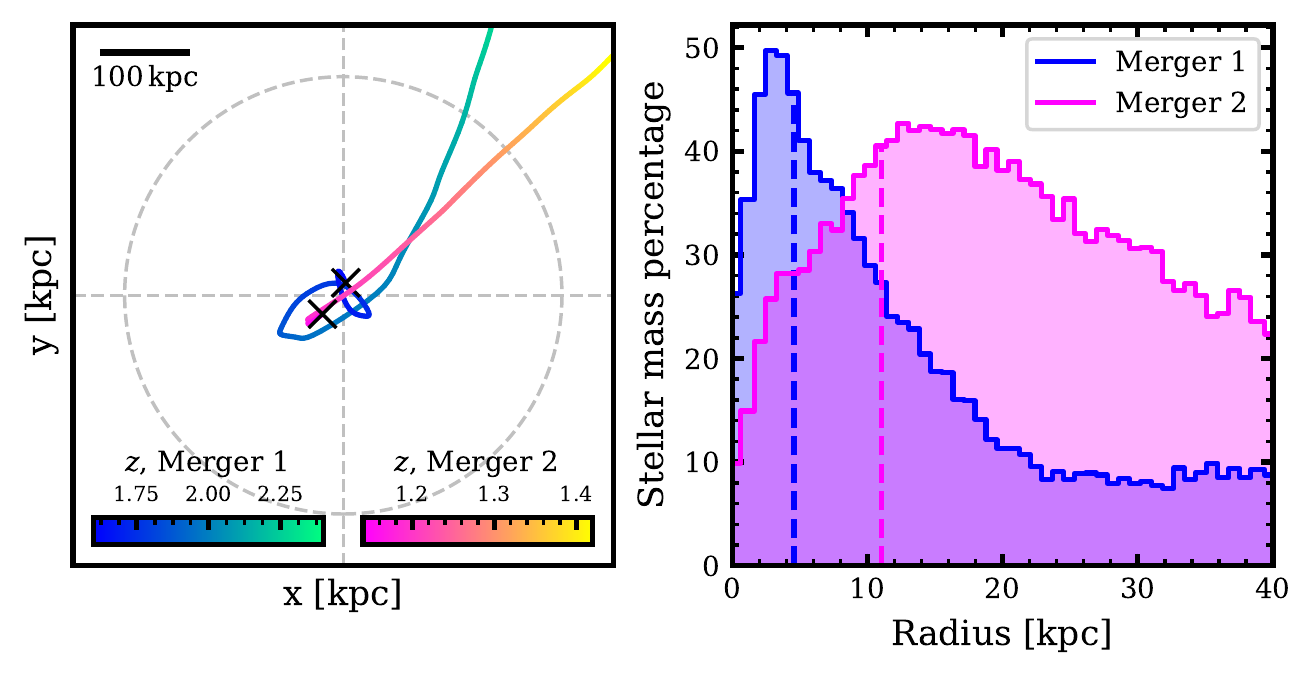}\\
\caption{\textit{Left panel:} The infall trajectories of Merger~1 and 2 (fit with a cubic spline) coloured by redshift. The orientation has been aligned on the angular momentum of the central host stars at $z=0$, such that the disc is in the $x-y$ plane. A grey dashed circle marks the $R_{200}$ virial radius of the host galaxy at $z=0$. Black crosses mark when the satellites can no longer be reliably tracked by {\sc subfind}.
\textit{Right panel:} The stellar mass contribution of each merger at $z=0$, as a percentage of all accreted stars. The peaks of the underlying stellar mass profiles are marked with vertical dashed lines. These do not necessarily align with the peaks of the accreted mass percentage.}
\label{fig:merger_illustration}
\end{figure}

We list properties for the two galaxies which we identify as being Kraken-like and GSE-like, hereafter Merger~1 and Merger~2, in Table \ref{tab:merger_comparison}. Au-24 has a chemically distinct disc with a dichotomy between the thin and thick discs \citep{verma2021}, with the thin disc developing some time after Merger~1. The possible accretion of the GSE is predicted to have occurred between $2\lesssim z \lesssim1.5$, with a stellar mass of order $\sim10^9\,$M$_{\odot}$ \citep{helmi2018, mackereth2019, Kruijssen2019, Naidu2022b}. The accretion of Kraken is more speculative, but \citet{Kruijssen2019} use simulations to estimate an accretion time before $z=2$, with a stellar mass of $\sim10^{8-9}\,$M$_{\odot}$. The mergers we identify in Au-24 are of a slightly higher mass, and accrete at a slightly later time, but are qualitatively similar. \par

We illustrate the trajectories of the two merging satellites with respect to the host galaxy in the left panel of Figure \ref{fig:merger_illustration}, aligned on the angular momentum of the central ($<0.1\,R_{200}$) stars at $z=0$. Both galaxies originate from similar locations in the cosmic web, but arrive at different epochs (see also Table \ref{tab:merger_comparison}). They infall in the plane of the disc at $z=0$, and this is a consequence of the disc reorienting itself to align with the incoming galaxies, a behaviour described in \citet{Auriga_discs}. In the right panel, we show the contribution of each merging galaxy to the total accreted stellar mass with radius (where accreted stars are defined as in \citealt{fattahi2019}). This shows that the accreted mass percentage from Merger~1 peaks around $3\,$kpc, whereas Merger~2 contributes a larger overall mass but only overtakes Merger~1 outside of $8\,$kpc. These satellites are the two single largest contributors to the accreted stars within $10\,$kpc at $z=0$. \par

We include a \href{https://www.youtube.com/watch?v=8HZLOUUSpFA}{movie} of the merger events as supplementary material, created using {\sc py-sphviewer} \citep{sphviewer}. \par

\subsection{Kinematics} \label{Kinematics}

\begin{figure}
\centering
  \setlength\tabcolsep{2pt}%
    \includegraphics[keepaspectratio, trim={0cm 0.1cm 0cm 0.1cm}, width=\columnwidth]{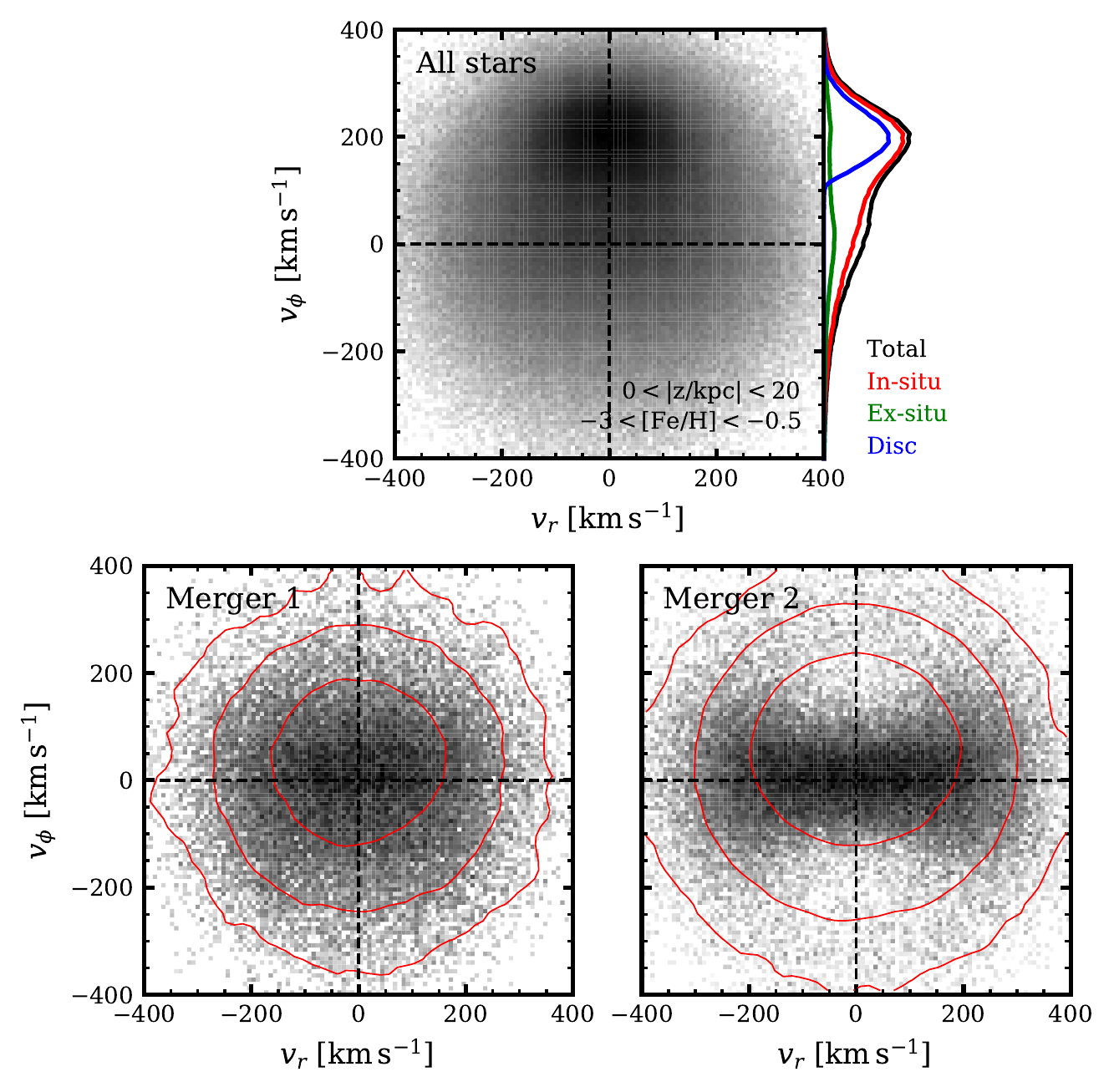}\\
\caption{\textit{Upper panel:} The logarithmic density of stars at $z=0$ in spherical $v_r, v_{\phi}$ coordinates, where the galaxy has been aligned on the angular momentum of the central host stars. We perform cuts on height above the disc and metallicity (indicated in the lower right-hand corner). The distributions plotted on the right-hand side represent the contribution of various components. We determine disc stars using the `circularity parameter', following the method and \citet{Auriga_discs}.
\textit{Lower panels:} Identical to the upper panel, except only including stars associated with Merger~1 and 2. Red contours mark the in-situ host stars present in the galaxy before each merger.}
\label{fig:velocity_space}
\end{figure}

We present velocity ellipsoid diagrams for metal-poor Au-24 stars in Figure \ref{fig:velocity_space}. In the upper panel, we show all stars where $R_{\rm G}<30\,$kpc and $<20\,$kpc above or below the disc at $z=0$. There is a feature corresponding to a co-rotating disc with a net rotational velocity of $200\,\text{km}\,\text{s}^{-1}$. In the lower panels, we show stars belonging to the merging haloes. Merger~1 appears as a near-isotropic distribution with a negligible net rotation. The configuration bears no substantial difference to the halo stars that had formed before the merger (red contours). Merger~2 is more radially anisotropic, which is reminiscent of the GSE as reported in \citet{belokurov2018}, with a small net rotation of $11\,\text{km}\,\text{s}^{-1}$. \par

\begin{figure}
\centering
  \setlength\tabcolsep{2pt}%
    \includegraphics[keepaspectratio, trim={0.3cm 0.1cm 0.1cm 0cm}, width=\columnwidth]{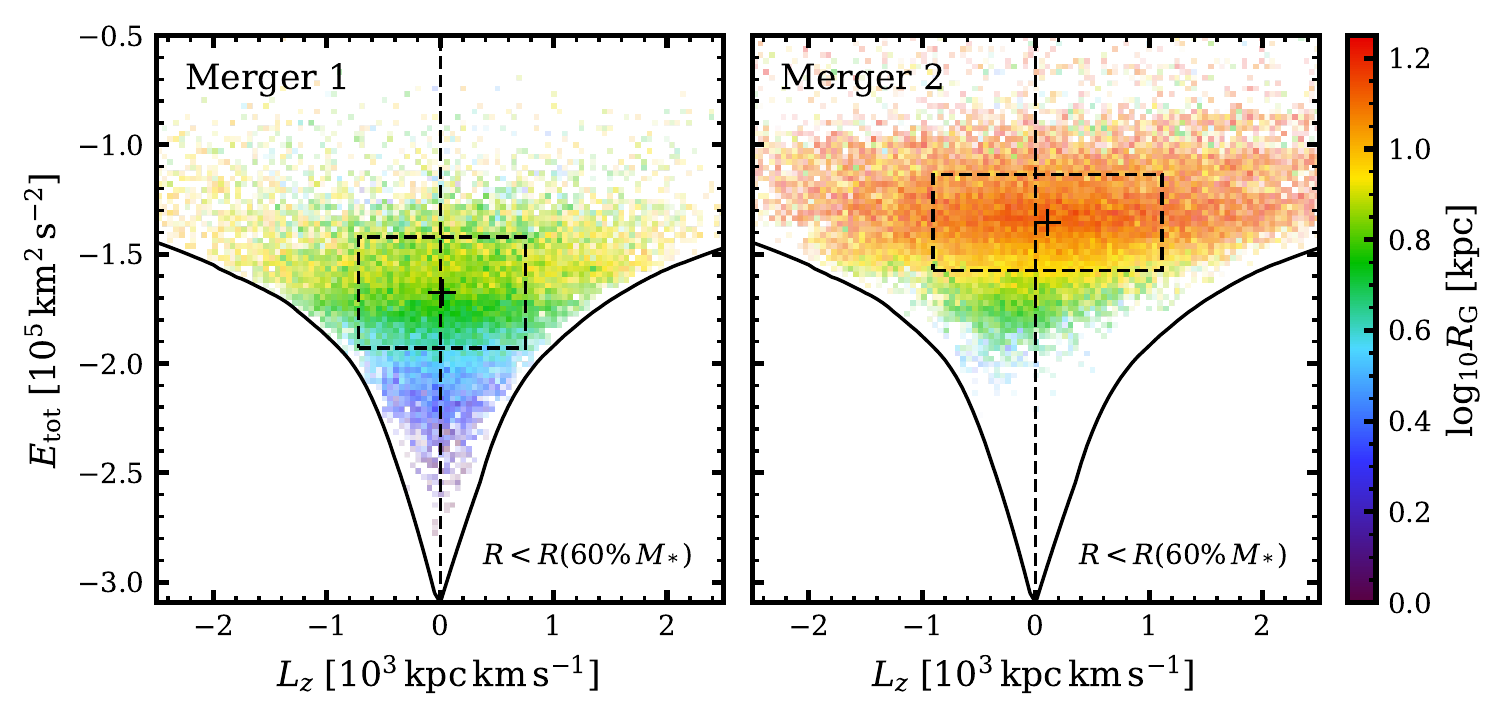}\\
\caption{The logarithmic density of stars associated with Merger~1 and 2 at $z=0$, shown in specific energy and the $z$-component of the angular momentum, where the galaxy has been aligned on the angular momentum of the central host stars. A colourmap has been overlayed, corresponding to the mean orbital radius in each bin. A radial cut has been performed, which excludes stars outside a radius containing 60 per cent of the total merger stellar mass. Black crosses indicate the mean of the distribution, with the dashed rectangle indicating the standard deviation.}
\label{fig:energy_space}
\end{figure}

Total local stellar energy versus angular momentum along the $z$-axis is shown for each merging galaxy in Figure \ref{fig:energy_space}. Colours correspond to the mean orbital radius of stars within each bin, showing that the stellar orbital energy is well correlated with orbital radius. To improve clarity, we consider radial cuts that encompass 60 per cent of the stellar mass within each merger, centred on the peak of the mass distribution (as shown in the right panel of Figure \ref{fig:merger_illustration}). The two mergers are roughly consistent with zero net angular momentum. Merger~1 occupies a lower energy level nearer the bottom of the potential well. The mean specific energies are found to be $-1.65\pm0.26\,\times10^5\,\text{km}^2\,\text{s}^{-2}$ for Merger~1 and $-1.33\pm0.22\,\times10^5\,\text{km}^2\,\text{s}^{-2}$ for Merger~2, yielding a difference of $\Delta E=0.32\,\times10^5\,\text{km}^2\,\text{s}^{-2}$. This is similar to the separation shown in Figure 2 of \citet{naidu2022} for observations of the MW, though in that case the stars attributed to Kraken and the GSE are systematically nearer the bottom of the potential well. \par

\subsection{Star formation history} \label{starformation}

Recently, \citet{ruiz2020} showed that repeated perturbations from the orbit of the Sagittarius dwarf galaxy left measurable imprints on the star formation history of the MW. Likewise, \citet{carme2019} found a burst of star formation coincident with the expected accretion time of the GSE. 
Merger-induced star formation also occurs in simulations \citep{tissera2002, bignone2019}, including Auriga galaxies \citep{gargiulo2019, grand2020}, with the cause attributed to gas infall driven by tidal forces, gas compression and/or shocking. \par

\begin{figure}
\centering
  \setlength\tabcolsep{2pt}%
    \includegraphics[keepaspectratio, trim={0cm 0.1cm 0cm 0.2cm}, width=\columnwidth]{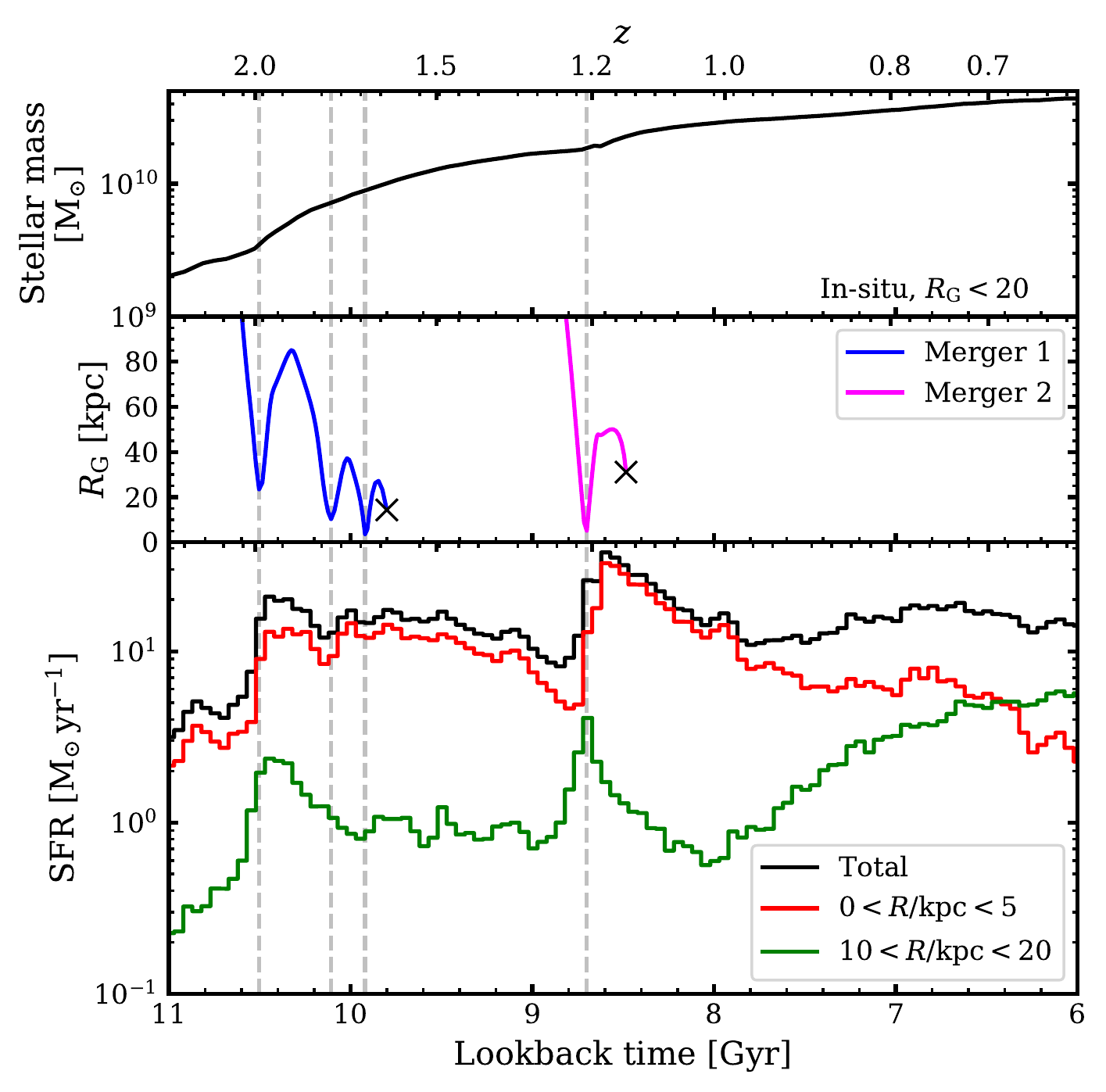}\\
\caption{\textit{Upper panel:} The stellar mass growth of in-situ stars within $20\,$kpc of the galactic centre. \textit{Middle panel:} The orbital radii of Merger~1 and 2. Black crosses mark when the satellites can no longer be reliably tracked by {\sc subfind}. Dashed grey lines mark each pericentre passage. \textit{Lower panel:} The mass-weighted star formation history of stars formed in-situ to the main host galaxy, in bins of $50\,$Myr, using the stellar birth masses. In addition, the SFR within two different radial bins are shown with red and green lines.}
\label{fig:SFH}
\end{figure}

Motivated by this, we show the star formation history of Au-24 for in-situ stars in several radial bins in the lower panel of Figure \ref{fig:SFH}. The middle panel shows the galacto-centric orbital radius of Merger~1 and 2 over the same period of time (fit with a cubic spline). The pericentre passes have been marked with dashed grey lines, which align with bursts of star formation in Au-24. The strong correlation suggests that the close passages of each merging galaxy are responsible for triggering renewed star formation. These bursts increase the star formation rate (SFR) by almost an order of magnitude, and persist for a short time after the merging satellite has disintegrated. We include the in-situ stellar mass growth in the upper panel. This shows sudden jumps associated with each merger event, and also highlights the low mass of the host galaxy prior to Merger~1. \par

We calculate the mass of stars formed during these two events by comparing the stellar mass growth over the redshift ranges $2.0>z>1.5$ and $1.2>z>1.0$, for all stars within $20\,$kpc at $z=0$. This yields $8.3\times10^9\,$M$_{\odot}$ and $9.4\times10^9\,$M$_{\odot}$, respectively. We then subtract a rough estimate for the mass of stars that would have formed if the mass growth continued unperturbed, finding new values of $5.7\times10^9\,$M$_{\odot}$ and $5.9\times10^9\,$M$_{\odot}$, respectively. This is several times the mass of stars donated by each merging galaxy at infall ($\sim4.3\times$ for Merger~1 and $\sim2.4\times$ for Merger~2).  \par

As shown in Table \ref{tab:merger_comparison}, both galaxies have a high gas fraction upon infall. The gas content is important not only for providing shocks to the in-situ gas, but also in supplying fuel for further star formation. The Auriga simulation analysed here includes tracer particles that follow the gas flow, and have a chance to be incorporated into any star particles that may form (see \citealt{grand2020} for a full description). We corral all stellar particles formed during the two previously defined starburst periods, and then select all those which inherited tracers. Then, we find all tracers that were ever in gas cells bound to each of Merger~1 and 2. This gives us a statistical approximation for the proportion of in-situ stars that formed directly from accreted gas. We find that 18 per cent of stars formed at $2.0>z>1.5$ came from gas associated with Merger~1, and 24 per cent of stars formed at $1.2>z>1.0$ came from gas associated with Merger~2. If we consider only the mass of in-situ stars formed with this accreted gas, it remains within a factor 2 of the peak stellar mass accreted directly from each merging galaxy (shown in Table \ref{tab:merger_comparison}). \par

\subsection{Chemistry} \label{sec:chemistry}

We also investigate the chemistry of stars belonging to Merger~1 and 2, comparing their iron ([Fe/H]) and $\alpha$-element ([Mg/Fe]) abundances. These values, calculated at $z=0$ and normalised to solar values following \citet{asplund2009}, are reported in Table \ref{tab:merger_comparison}. The abundance differences between the two galaxies are small, and would be difficult to confirm observationally. \par

\begin{figure}
\centering
  \setlength\tabcolsep{2pt}%
    \includegraphics[keepaspectratio, trim={0.0cm 0.1cm 0.0cm 0.2cm}, width=\columnwidth]{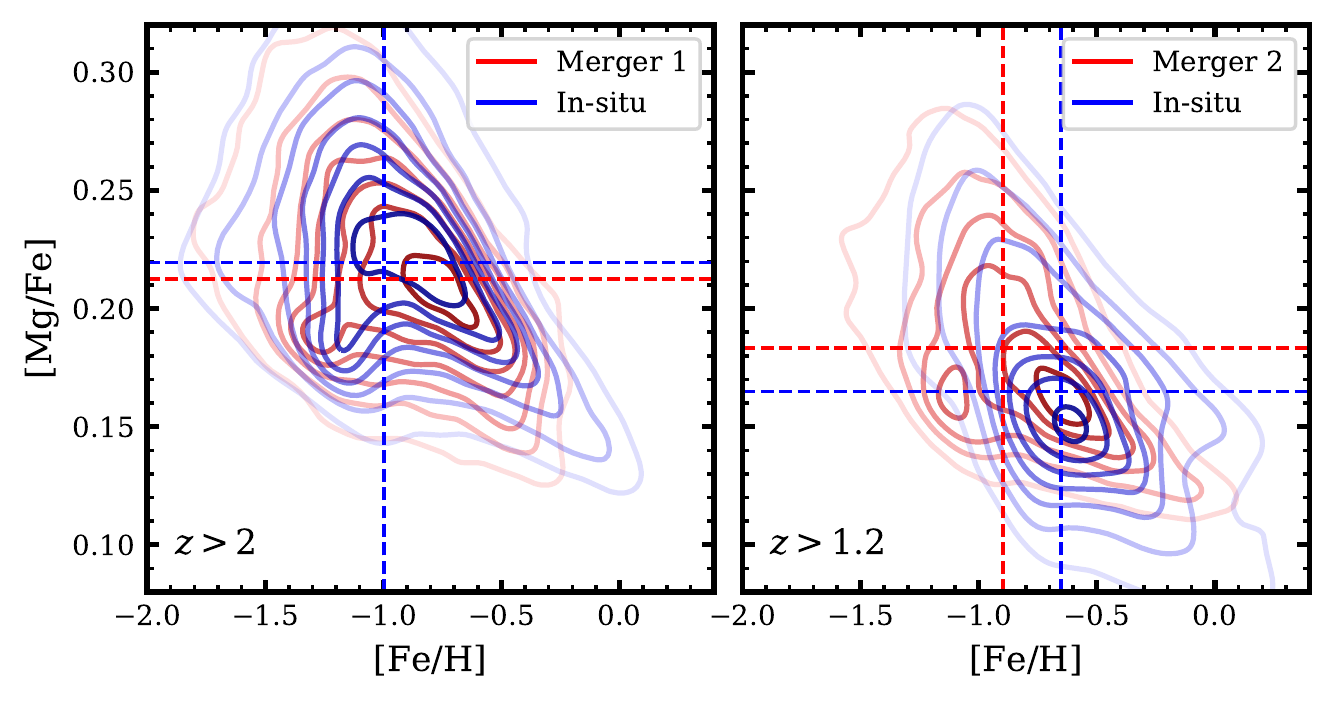}\\
\caption{A comparison between stellar metallicities of the in-situ population and stars associated with each merging galaxy, where metallicities are normalised as in \citet{grand2020}. We consider only stars born before the first pericentre passage (indicated in the bottom-left corner). Contours denote the extent of the distributions, but are not weighted proportionally to each other. Dashed lines mark the medians.}
\label{fig:chemsitry}
\end{figure}

We compare the chemistry of merger and in-situ stars prior to the first pericentre passage of each merging galaxy in Figure \ref{fig:chemsitry}. The chemistry is remarkably alike in the case of Merger~1, which is unsurprising given the high stellar mass merger ratio (see Table \ref{tab:merger_comparison}). Ancient galaxies of increasingly similar mass should be at an increasingly similar evolutionary phase. We find similar results when comparing the abundances of other metal species, including heavy elements. As a result, the accreted gas and stars do not directly alter the chemical evolution within the host. \par

The chemistry in the case of Merger~2 is more distinguishable, with the median [Fe/H] differing by 0.25 from that of in-situ stars, though the accreted and in-situ distributions have substantial overlap. Observational surveys show that the evolutionary track of the real GSE is offset towards much lower [Mg/Fe] abundances than the MW \citep[e.g.][Fig. 6]{horta2022}, and we stress that our analogues are intended for qualitative and not direct comparison. \par

Recently, \citet{sergey2022} suggest that the differences between abundance ratios of in-situ and accreted stars could help constrain the accretion history of the MW. These differences are most pronounced between the highest metallicity stars. However, if the elemental abundances of host and satellite are near-identical at the time of merging (as in our model), then an independent method would still be required to disentangle merger from in-situ stars. As such, these tests cannot be performed with real data.

Assuming that such distinction between high-$z$ accreted and in-situ stars were possible, it should be noted that the fraction of Merger~1 stars in the highest-metallicity bins ($-0.5 <[\rm{Fe/H}]< 0$ and $0 < [\rm{Fe/H}]< 0.5$, where the larger differences are advocated), correspond to about 11 per cent to 0.01 per cent, or about 1 star in a million if one were to test this. \par

\section{Discussion} \label{sec:discussion}

As shown in Section \ref{sec:chemistry}, the chemistry of stars belonging to Merger~1 are difficult to distinguish from in-situ stars at the time of accretion -- and this is a direct consequence the two galaxies being of similar mass, in other words at a similar stage in their chemical evolution history. However, stars attributed to `Kraken' are frequently selected with the aid of chemical cuts. \par

In \citet{horta2021}, Kraken stars are selected using cuts in the [Mg/Mn] versus [Al/Fe] chemical plane as well as energy space. These cuts yield characteristic metallicities in ([Fe/H], [Mg/Fe]) of $(-1.25, 0.3)$.
Using data from APOGEE, \citet{belokurov2022} identify a highly pure population of old, in-situ stars which they name \textit{Aurora}. There are many of these \textit{Aurora} stars with [Mg/Mn] and [Al/Fe] abundance ratios which are a good match for the aforementioned characteristic [Fe/H] and [Mg/Fe] metallicities. From inspection of Figure 2 in \citet{belokurov2022}, this is likely due to contamination from in-situ high-$\alpha$ stars. Indeed, \citet{horta2021} acknowledge such stars as a potential source of contamination.
To expand upon this, Figure 13 of \citet{wheeler2020} includes chemical abundances of the MW from the LaMOST survey. The known values in the $\alpha$-rich thick disc are [Mg/Fe]$\approx0.25$ and [Mn/Fe]$\approx-0.2$, leading to [Mn/Mg]$\approx0.45$ -- once more compatible with the values used to select Kraken stars in \citet{horta2021}. Similarly, there are values of [Eu/Fe]$\approx0.25$ leading to [Eu/Mg]$\approx0$, comparing well with [Eu/Mg]$\approx-0.1$ for stars identified in \citet{naidu2022} attributed to belong to Kraken. Abundances can also be derived for [Al/Fe] and [Mg/Fe], leading to similar conclusions. \par

Therefore, stars attributed to Kraken may in fact be in-situ \textit{Aurora} population stars (and also unlikely to originate from the GSE debris because we would then expect to observe an excess of metal-rich stars towards the bulge, \citealt{amarante2022}). Such an interpretation is also supported by the recent stellar halo study of \citet{myeong2022}. This should serve as caution against using chemical cuts to differentiate massive mergers at high redshift, and the need to devise alternative tests and falsifiable predictions for such scenarios. \par

\section{Conclusions} \label{sec:conclusion}

We have identified an Auriga MW-analogue with an early accretion history dominated by two massive satellites, which we call Merger~1 and 2. These galaxies bear a strong qualitative resemblance to the expected properties of Kraken and the GSE, in terms of their infall times and masses, the kinematics of their debris, and overall chemical trends. Our key results are as follows:

\begin{itemize}
\setlength{\itemindent}{0.5em}
    \item The stellar debris from Merger~2 adopts an elongated shape in spherical $v_r, v_{\phi}$ coordinates, likely due to its radial infall. This contrasts with the isotropic velocity ellipsoid of Merger~1, which conforms to in-situ halo stars.
    \item Merger~1, being a high mass ratio merger, is at a similar stage in its chemical evolution to the host galaxy at the time of merging. This makes the two indistinguishable in chemical abundance space, and casts doubt that evidence of the merger debris could be obtained from chemistry and dynamics alone.
    \item Although there is no certain evidence for Kraken in current chemical abundance observations, our model produces abundance patterns consistent with observations of the MW. Therefore, the presence of a Kraken-like merger in the MW cannot be ruled out.
    \item The overall in-situ SFR at the centre of the host galaxy is excited by the two massive mergers. The excess stellar mass formed during these times exceeds that of the accreted stellar mass by a factor of $\sim2$-$5$. Such pronounced bursts of star formation may be detectable. 
\end{itemize}

Contrary to the assertion that the Kraken debris can be distinguished from its chemodynamics, we propose a falsifiable prediction that if the accretion history of the MW is dominated by two massive mergers, then the stars towards its centre will have a dual-peaked star formation history. This may be revealed with future spectroscopic campaigns directed towards the MW bulge or even around the solar neighbourhood (i.e. MOONS \citealp{moons2020}, SDSS-V \citealp{sdss52017}, 4MOST \citealp{4MOST}, WEAVE \citealp{WEAVE}). However, the star formation history would need to be sufficiently well-resolved to differentiate multiple SFR peaks. While this could prove challenging for such high lookback times and if the two mergers occurred within a relatively close time-frame, it is within the reach and science goals of upcoming and future asteroseismic missions such as PLATO \citep{Miglio2017} and HAYDN \citep{miglio2021}. \par

\section*{Acknowledgements}
CL \& MO acknowledge funding from the European Research Council (ERC) under the European Union’s Horizon 2020 research and innovation programme (grant agreement No. 852839).
CL thanks Vasily Belokurov for enlightening discussions.
RG acknowledges financial support from the Spanish Ministry of Science and Innovation (MICINN) through the Spanish State Research Agency, under the Severo Ochoa Program 2020-2023 (CEX2019-000920-S).
FvdV is supported by a Royal Society University Research Fellowship.

\section*{Data Availability}

The data underlying this article will be shared on reasonable request to the corresponding author.


\bibliographystyle{mnras}
\bibliography{references} 







\bsp	
\label{lastpage}
\end{document}